\documentclass[10pt, journal]{IEEEtran}
\IEEEoverridecommandlockouts
\usepackage{cite}
\usepackage{amsmath,amssymb,amsfonts}
\usepackage{algorithmic}
\usepackage{graphicx}
\usepackage{textcomp}
\usepackage{xcolor}
\usepackage{epstopdf}
\usepackage[english]{babel}
\begin{document}
\title{\makebox[\linewidth]{\parbox{\dimexpr\textwidth+0cm\relax}{\centering Nonlinear Multi-Carrier System with Signal Clipping: Measurement, Analysis, and Optimization}}}
\author
{
Yuyang~Du,~\IEEEmembership{Graduate Student Member,~IEEE},~\
Liang~Hao,~\IEEEmembership{Member,~IEEE},~\
\\Yiming~Lei,~\IEEEmembership{Member,~IEEE},~\
Qun~Yang, ~\IEEEmembership{Student Member,~IEEE},~\
Shiqi~Xu,~\IEEEmembership{Student Member,~IEEE},~\
\thanks{Y. Du and Y. Lei are with the School of Electronics, Peking University, Beijing, China. H. Liang is with 2012 Laboratory, Huawei Technologies Co., Ltd, Beijing, China. Q. Yang is with the Department of Information Engineering, The Chinese University of Hong Kong, Shatian, Hong Kong SAR. S. Xu is with the School of Electronics and Information Technology, Sun Yat-sen University, Guangdong, China. An early version of this paper\cite{ref-VTC-conf} has been presented in IEEE VTC2023.}
\vspace{-2em}
}
\maketitle

\begin{abstract}
Signal clipping is a well-established method employed in orthogonal frequency division multiplexing (OFDM) systems to mitigate peak-to-average power ratio (PAPR). The utilization of this technique is widespread in electronic devices with limited power or resource capabilities due to its high efficiency and low complexity. While clipping effectively diminishes nonlinear distortion stemming from power amplifiers (PAs), it introduces additional distortion known as clipping distortion. The optimization of system performance, considering both clipping distortions and the nonlinearity of PAs, remains an unresolved challenge due to the intricate modeling of PAs. In this paper, we undertake an analysis of PA nonlinearity utilizing the Bessel-Fourier PA (BFPA) model and simplify its power expression through inter-modulation product (IMP) analysis. We mathematically derive expressions for the receiver signal-to-noise ratio (SNR) and system symbol error rate (SER) for nonlinear clipped OFDM systems. By means of these derivations, we explore the optimal system configuration required to achieve the lower bound of SER in practical OFDM systems, taking into account both PA nonlinearity and clipping distortion. The results and methodologies presented in this paper contribute to an improved comprehension of system-level optimization in nonlinear OFDM systems employing clipping technology.
\end{abstract}


\section{Introduction}\label{Sec-I}
Power amplifiers (PAs) exhibit nonlinear characteristics when faced with high-power input signals, which is commonly referred to as PA nonlinearity\cite{ref-VTC-conf, ref063a,ref063b,ref063c,ref063d,ref063e,ref063f,ref063g}. The rapid progress of multiple-input and multiple-output (MIMO) technology, coupled with the growing number of subcarriers, has resulted in a significant rise in the peak-to-average power ratio (PAPR) of orthogonal frequency division multiplexing (OFDM) devices. This increase in PAPR poses a major challenge in practical OFDM systems, as it can lead to reduced power efficiency, degraded system performance, and potential distortion in the transmitted signals. PA's nonlinearity has become a vital concern, as it may cause significant performance degradation when faced with high PAPR \cite{ref056,ref051}. To ensure system linearity, designers are compelled to reduce the input power of the amplifier, albeit at the cost of energy efficiency\cite{ref055}. Reducing PAPR, as a result, has emerged as a critical problem in OFDM systems and has attracted significant research attention.%\cite{ref059, ref060, ref061, ref062}.

Several established PAPR reduction algorithms \cite{ref041,ref003,ref023,ref024} have been extensively studied. However, these conventional algorithms, such as partial transmit sequence (PTS) and selected mapping (SLM), involve complex signal processing, making them unsuitable for resource-limited or battery-driven devices, which is common in sensors and Internet of Things networks. %\cite{ref000b}. 

Signal clipping offers a less complex PAPR reduction approach \cite{ref068a, ref068b, ref068c}. In OFDM systems, clipping is realized in the time-domain signals to limit their magnitude to a predetermined threshold. This approach effectively limits the PAPR of OFDM signals within an upper bound with relatively small computation requirements and is therefore very popular in resource-constrained OFDM devices \cite{ref045}. However, as a nonlinear operation, signal clipping introduces additional frequency components that never appear in the original signal, bringing both out-of-band radiation and in-band distortion to the original signal. Although radiation spilling beyond the target bandwidth can be restricted by filtering, how to deal with the in-band distortion is still a challenging problem. 

Prior research has made efforts to investigate and mitigate the in-band distortion caused by signal clipping. In \cite{ref008}, a novel Bayesian approach was implemented to recover the clipped signals of an OFDM system, aiming to minimize the impact of in-band distortion at the receiver side. Another technique, proposed in a separate publication \cite{ref009}, is known as the repeated clipping and filtering (RCF) method. The RCF technique focuses on reducing distortion on individual tones of the OFDM signal. Researchers have also explored optimization techniques to enhance the performance of signal clipping for PAPR reduction in OFDM systems. In a couple of studies \cite{ref010,ref011}, the authors optimized both the clipping and filtering stages to achieve improved PAPR reduction while simultaneously mitigating in-band distortion. Alternative approaches have been investigated to tackle the computational complexity associated with clipped OFDM systems. For example, \cite{ref012} applies convolutional neural networks (CNN) to reduce computational complexity while maintaining PAPR reduction performance. Additionally, compressed sensing techniques were employed in another study \cite{ref013} to recover the signal clipping noise in clipped OFDM signals.

However, the joint impact of PA nonlinearity and clipping distortion has received little attention in previous studies. Intuitively, the choice of clipping level introduces a trade-off between PA nonlinearity and clipping noise: a higher clipping level reduces PA nonlinearity but increases clipping noise, while lower clipping levels decrease clipping noise but amplify nonlinear distortion. A previous research \cite{ref067} tried to understand the problem via simulations. However, the complex settings in real-world engineering scenarios necessitate a \textit{theoretical} analysis of the trade-off between these two distortion sources. Otherwise, with simulations only, we are unable to comprehensively understand the problem, let alone optimize the system setting to achieve the best performance.

In light of this, the primary objective of this paper is to thoroughly investigate the trade-off between clipping distortion and PA nonlinearity in practical OFDM systems and make optimizations accordingly. By addressing this trade-off, we aim to enhance the overall system performance and mitigate the adverse effects of both clipping distortion and PA nonlinearity. The contributions of this research can be summarized as follows:
\begin{itemize}
\item To simplify the representation of PA nonlinearity, we utilize the Bessel-Fourier PA (BFPA) model. Through an analysis of the inter-modulation product (IMP) in the PA's output, we derive the signal-to-noise ratio (SNR) for the studied system, where both nonlinear PAs and signal clipping are taken into consideration.
\item Our optimization problem is approached from three distinct scenarios. In the first scenario, we consider a known power level for the PA's nonlinear distortion. By deriving the optimal clipping level, we aim to minimize the symbol error rate (SER). Moving on to the second scenario, we assume the power of clipping distortion is known. In this case, we derive the optimal operating point for the PA that minimizes the SER. Lastly, we address the scenario where both PA nonlinearity and clipping distortion are unknown variables. We prove the existence of a global minimum SER and derive the optimal signal clipping level and PA operating point. Besides, we obtain a closed-form expression for the global SER lower bound, which enhances our understanding of the system's performance.
\item We thoroughly examine the impact of various system parameters on both the SER and total degradation (TD) in the presence of signal clipping noise and PA nonlinearity. Our investigation offers a comprehensive analysis of how these system parameters contribute to the overall performance degradation.
\end{itemize}

This paper is organized as follows. Section II presents the framework of the studied nonlinear clipped OFDM system. Section III-A builds the BFPA model, Section III-B analyzes the PA nonlinear distortion, and Section III-C derives the closed-form expression of SER. Then in Section IV, SER optimizations with different constraints are presented in detail. Section V gives simulation results and explains our experimental observations. Section VI concludes this paper.

This paper uses the following notation conventions. we denote matrices by bold capital letters, e.g., $\mathbf{M}$. And we denote the element of a matrix and the conjugate of the element by ${{\left[ \mathbf{M} \right]}{i,j}}$ and $\left[ \mathbf{M} \right]{i,j}^{*}$, respectively. We represent the Frobenius norm of a matrix by ${{\left| \mathbf{H} \right|}_{F}}$. We denote the expectation operator and the variance operator by $E(\cdot)$ and $Var(\cdot)$, respectively. Furthermore, $\mathsf{\mathcal{C}\mathcal{N}}(m, {\sigma }^{2})$ represents a complex Gaussian random variable with a mean of $m$ and a variance of ${{\sigma }^{2}}$.

\section{System Model}\label{Sec-II}
We first present the general framework of the studied system in Fig. 1. The number of subcarriers is denoted as $N_{S}$. Additionally, the system we considered is comprised of $N_{R}$ receive antennas and $N_{T}$ transmit antennas. We then represent the source bit stream by vector $\mathbf{U}$ and denote the output of the $M$-order Quadrature Amplitude Modulation (M-QAM) modulator by matrix $\mathbf{M}$. Further, the modulated symbol transmitted through the $t^{th}$ transmitter antenna (where $t=1,2,...,N_T$) during the $s^{th}$ time slot (where $s=1,2,...,N_S$) is denoted as $\left[ \mathbf{M} \right]_{t,s}$.
\begin{figure}[htbp]
    \label{figure_1}
  \centering
  \includegraphics[width=0.4\textwidth]{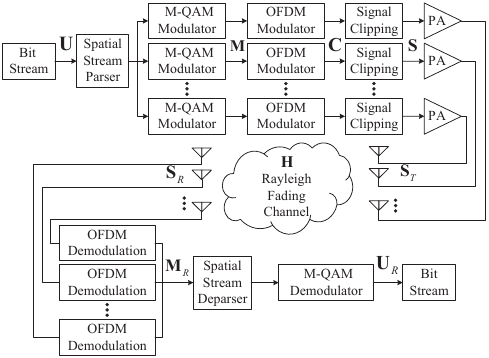}\\
  \caption{The framework of the investigated OFDM system incorporates both a nonlinear PA and signal clipping.}
  \label{SystemModel}
\end{figure}

We make an assumption that the inverse fast Fourier transform (IFFT) module has a length $N_{S}$, which is equal to the number of subcarriers. Consequently, the signal obtained after the OFDM modulator can be expressed as
\begin{equation}
\mathbf{C}=\mathbf{M}\mathbf{E}^{+}
\end{equation}
where $\mathbf{E}^{+}$ represents the IFFT operation within the OFDM modulation. The $(k,s)$ element within the matrix can be written as \cite{ref065}
\begin{equation}
{\left[ \mathbf{E} \right]}^{+}_{k,s}=e^{j2\pi(s-1)(k-1)/N_S}
\end{equation}

We define the system input power as $E_{U}$, which represents the system operating point. Additionally, we denote the input power of the signal clipping blocks as $E_{C}$. Furthermore, at the $t^{th}$ transmit antenna, the $s^{th}$ post-IFFT symbol transmitted is denoted by ${\left[ \mathbf{C} \right]}_{t,s}$. The expression for $E_U$ can be written as
\begin{equation}\label{eq0001}
{E_U} {=} \frac{\sum\limits_{x = 1}^{{N_T}{N_S}} {E( {{{\left[ {\bf{U}} \right]}_x}\left[ {\bf{U}} \right]_x^*} )}}{{{N_S}}}
 {=} \frac{\sum\limits_{t = 1}^{{N_T}} {\sum\limits_{s = 1}^{{N_S}} {E ( {{{\left[ {\bf{C}} \right]}_{t,s}}\left[ {\bf{C}} \right]_{t,s}^*} )} }}{{{N_S}}}
\end{equation}

We then express the output signal of the clipping module as
\begin{equation}\label{eq0002}
\left[\mathbf{S} \right]_{t,s}=\left\{ \begin{aligned}
  &  \left[\mathbf{C} \right]_{t,s} \text{,             when  }{\left[\mathbf{C} \right]_{t,s}}\le \eta \sqrt{{{E}_{U}}} \\
 & \eta \sqrt{{{E}_{U}}}e^{j\angle{\left[\mathbf{C} \right]_{t,s}}},\text{  otherwise} \\
\end{aligned} \right.
\end{equation}
where the phase of the transmitted symbol is denoted by $\angle (.)$, and the signal clipping level is represented by $\eta$. From (\ref{eq0002}), we see that the output power of the clipped signal is upper-bounded by $\eta^{2}E_{U}$.

Building upon Equation (\ref{eq0002}), the scaling factor $\beta$ for signal power is written as,
\begin{equation}\label{eq0003}
\beta=\eta \int_{\eta}^{\infty}e^{-t^{2}}dt + 1 -e^{-\eta^{2}}
\end{equation}

We then write the signal matrix after clipping as
\begin{equation}\label{eq0004}
\mathbf{S} = \mathbf{D}_{clip} + \beta \mathbf{C}
\end{equation}
where  ${{\mathbf{D}}_{clip}}$ denotes the signal clipping distortion, and we have ${{\mathbf{D}}_{clip}}\sim G\left( 0,{{D}_{k}}{{E}_{U}} \right)$.

We write the coefficient of clipping distortion power as
\begin{equation}\label{eq0005}
\begin{aligned}
 {{D}_{k}}=\frac{1}{J}FFT{{\left\{ \mathbf{S} \right\}}_{k}} -{{\beta}^{2}}/{{\left( 1-{{e}^{-{{\eta }^{2}}}} \right)}}
\end{aligned}
\end{equation}
where the over-sampling factor is written as $J$, and the $k^{th}$ output of the $JN$-point fast Fourier transform (FFT) is denoted by $FFT\left\{ \right\}_{k}$.

The power of the signal clipping module's output can be expressed as
\begin{equation}\label{eq0006}
{E_S} = \frac{1}{{{N_S}}}\sum\limits_{t = 1}^{{N_T}} {\sum\limits_{s = 1}^{{N_S}} {E\left( {{{\left[ {\bf{S}} \right]}_{t,s}}\left[ {\bf{S}} \right]_{t,s}^*} \right)} }
=D_{k}E_{U} + \beta^{2}E_{U}
\end{equation}

As in Fig. 1, we represent the input of nonlinear PAs by ${\mathbf{S}}$, which is the clipped output of prior modules. And we represent the output symbol of PA by $\mathbf{S}_T$, which is expressed as
\begin{equation}\label{eq0007}
\mathbf{S}_{T}=\alpha\mathbf{S}+\mathbf{D}_{non}=\alpha\beta\mathbf{C}+\alpha \mathbf{D}_{clip}+\mathbf{D}_{non}
\end{equation}
where $\mathbf{D}_{non}$ represents the nonlinear distortion of amplifiers and $\alpha$ denotes the PA's linear gain factor.

Radio-frequency signals received at the receiver antenna can be written as
\begin{equation}\label{eq0008}
{{\mathbf{S}}_{R}}=\mathbf{H}{{\mathbf{S}}_{T}}+{{\mathbf{D}}_{ch}}
\end{equation}
where the channel noise is represented by $\mathbf{D}_{ch}$, and the uncorrelated elements of $\mathbf{D}_{ch}$ can be written as $\mathsf{\mathcal{C}\mathcal{N}}(0,\sigma_{ch}^{2})$. Further, the quasi-static Rayleigh channel applied in the system is denoted by $\mathbf{H}$.

The output signal of the FFT module (i.e., the OFDM demodulator) is expressed as
\begin{equation}\label{eq0009}
\mathbf{M}_{R}={\mathbf{S}}_{R}\mathbf{E}^{-}=(\alpha\mathbf{HS}+\mathbf{HD}_{non}+\mathbf{D}_{ch})\mathbf{E}^{-}
\end{equation}
where the $s^{th}$ post-FFT symbol obtained through the $r^{th} \left( r=1,2,...,N_R \right)$ receiver antenna is denoted by $\left[ \mathbf{M_R} \right]_{r,s}$, and we know
\begin{equation}
\left[ \mathbf{E^-} \right]_{k,s} = {N_S^{-1}}e^{-j2\pi(s-1)(k-1)/{N_S}}
\end{equation}

As in \cite{ref058}, we model the PA's nonlinear in-band distortion $\mathbf{D}_{non}\mathbf{E}^-$ as a Gaussian signal. The output of the M-QAM demodulator can be written as
\begin{equation}\label{eq0010}
{{\mathbf{U}}_{R}}= \mathbf{W} + \alpha\beta\mathbf{H}\mathbf{U}
\end{equation}
where the sum of system impairments is denoted by $\mathbf{W}$, and the output symbol of the source bit stream is written as $\mathbf{U}$.

We write $\mathbf{W}$ as
\begin{equation}\label{eq0011}
\mathbf{W} = (\alpha\mathbf{H}{{\mathbf{D}}_{clip}}+\mathbf{H}{{\mathbf{D}}_{non}}+{{\mathbf{D}}_{ch}})\mathbf{E}^{-}
\end{equation}

\section{BFPA Model and SNE/SER Analysis}\label{Sec-III}
In subsection A, we present how we measure a PA chip and give the measurement results for building the PA model. And then in subsection B, we use the BA model to simplify the SNR analysis. Finally, we derive the SER performance in subsection C.

\subsection{PA modeling and measurements}
Numerous prior studies have extensively explored analytical models for nonlinear PAs in the literature. The power spectrum of PA's nonlinear distortion has been investigated in prior works like \cite{ref028}. These works also established the relationship between the PA operating point and the signal SNR at the receiver antennas. However, they often involve complex calculations, including integrals and partial differentials, which hinder their practical applicability for extensive analysis of nonlinear PA systems. Some works also develop simple PA behavior models with consideration of memory effect \cite{ref069a, ref069b, ref069c}. However, the primary focus of this paper is to understand the impact of PA nonlinearity and clipping distortion. We are not interested in considering the memory effect, which brings additional computation complexity.

This paper utilizes a simple memoryless PA model known as the "BFPA model" to capture the PA nonlinearity. The concept of BFPA was developed in \cite{ref019}, comprehensively analyzed in \cite{ref052}, and then became widely used in later work like\cite{ref056,ref065, ref057, ref066, ref058}. With the BFPA model, we focus on the analysis of the amplitude modulation to amplitude modulation (AM/AM) characteristic of the PA. By investigating this characteristic, we aim to understand the nonlinear behavior exhibited by the PA and its impact on system performance. The equivalent PA model is derived based on memoryless characteristics extracted from extensive laboratory measurements of the input-output behaviors of a commercially available PA. Fig. 2 presents the experimental setup employed for the PA measurement in this research.
\begin{figure}[htbp]
  \centering
  \includegraphics[width=0.4\textwidth]{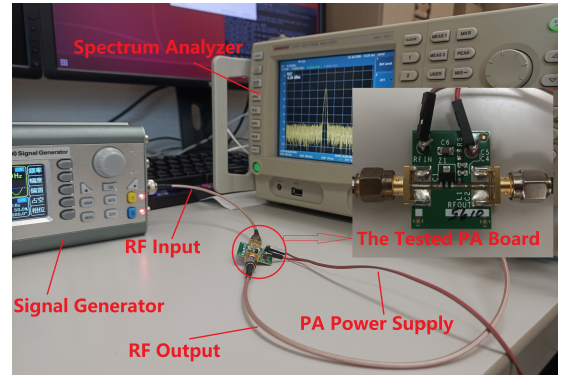}\\
  \caption{Experimental setup for the PA measurement.}\label{PA_Measurement}
\end{figure}

We now explain our measurement as follows. We first generate a 2.4GHz single-tone signal with the signal generator. We then transmit the signal via the test PA and measure the PA's output with a spectrum analyzer. We keep testing different signal power and recording the PA's output power. Finally, we plot the scatter chart as in Fig. 3 and fit the AM-AM curve for the tested PA. For better illustration, we also present the ideal AM-AM curve for the tested PA by assuming that the PA is a linear one.
\begin{figure}[htbp]
  \centering
  \includegraphics[width=0.4\textwidth]{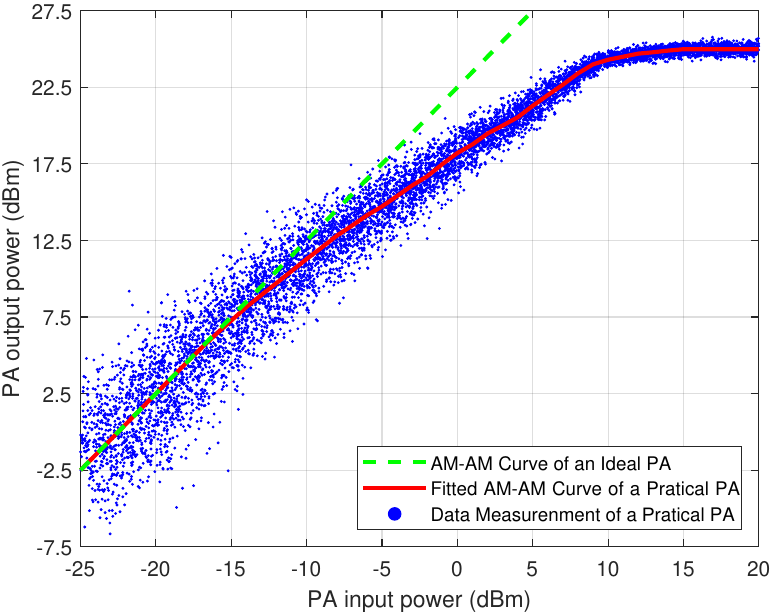}\\
  \caption{The scatter chart of the measured PA and the fitted AM-AM curve of the PA.}\label{AM_AM}
\end{figure}

After the measurement, we use the fitted AM-AM curve to build the BFPA model as in \cite{ref052}. Given a $P$-order BFPA model, let us denote the $p^{th}$ order coefficients by $b_p$ and denote the dynamic range of the model by $P_{mod}$.

\subsection{Polynomial Function of Receive SNR}
Upon examination of the right side of (\ref{eq0009}), it is evident that the power of the desired signal can be further written as ${\alpha^{2} \beta^{2}\left| \mathbf{H} \right|{F}^{2}E_U}$. Additionally, the power of the impairment symbol $\mathbf{W}$ is represented by $E_W$. Consequently, the SNR at the receiver, following the OFDM demodulator, can be represented as
\begin{equation}\label{eq0012}
\gamma =\frac{\alpha^{2} \beta^{2}\left\| \mathbf{H} \right\|_{F}^{2}E_U}{E_W}
\end{equation}
where $E_{{W}}$ encompasses various sources of system distortion, including channel noise, signal clipping noise, and PA nonlinearity. In light of these considerations, we express it as
\begin{equation}\label{eq0013}
\begin{aligned}
&E_{W} =  \sigma _{ch}^{2} +  {{\alpha }^{2}}{{D}_{k}}{{E}_{U}} \\
&\text{ }+ \frac{1}{{{N}_{S}}} \sum\limits_{t=1}^{{{N}_{T}}}{\sum\limits_{s=1}^{{{N}_{S}}}{E\left( {{\left[ \mathbf{H}{{\mathbf{D}}_{non}}{{\mathbf{E}}^{-}} \right]}_{t,s}}\cdot \left[ \mathbf{H}{{\mathbf{D}}_{non}}{{\mathbf{E}}^{-}} \right]_{t,s}^{*} \right)}} \\
\end{aligned}
\end{equation}

[R3-5]We now delve into the modeling of PA nonlinearity to simplify the last term in (\ref{eq0013}). In this paper, we follow the IMP analysis presented in our previous paper \cite{ref066} to approximate the power of PA's nonlinear distortion. Due to space limits, this paper does not repeat these derivation details. Instead, we present the outline of the derivation and refer our reader to \cite{ref066} for detailed analysis and discussions. In general, the last term in (\ref{eq0013}) can be expressed as
\begin{equation}\label{eq0029}
\sum\limits_{\delta=3,5...}
\sum\limits_{s = 1}^{{N_S}} {\phi(\delta,s)}   \Omega  \left( {\delta,{\beta ^2}{E_U}} \right)
\end{equation}
where $\phi(\delta,s)$ denotes the number of $\delta$ order IMPs falling at the $s^{th}$ OFDM subcarrier, and it can be counted through a numerical process \cite{ref057}. Additionally, $\Omega  (\eta,\beta^{2}E_{U})$ is the power of an individual $\delta$ order IMP, and it can be further written as
\begin{equation}\label{eq0022}
\Omega \left( \delta ,\beta^{2}{{E}_{U}} \right)={{\left| \sum\limits_{p=1}^{P}{\left[ {{b}_{p}}\cdot J_{0}^{{{N}_{S}}-\delta }\left( V_{p} \right)\cdot J_{1}^{\delta }\left( V_{p} \right) \right]} \right|}^{2}}
\end{equation}
where $J_{g_{m}}$ denote the $g_m^{th}$ order Bessel function of the first kind, and $b_p$ denotes the coefficients of a $P$-order BFPA model. Furthermore, $V_p$ can be written as
\begin{equation}\label{eq0023}
V_{p}=\frac{2p\pi}{{P_{mod}}\sqrt{{\beta^{2}E_{U}}/{N_{T}N_{R}}}}
\end{equation}
where $P_{mod}$ is the dynamic range of the $P$-order BFPA.

Further considering the range of the Bessel kernel and assuming $\delta = 1$, we have the power of a first-order IMP (i.e., the wanted signal) as
\begin{equation}\label{eq0024}
\begin{aligned}
&\Omega  \left( {1,{\beta ^2}{E_U}} \right)
= \frac{{{\beta ^2}{E_U}}}{{{N_T}{N_S}}}{\left| {\sum\limits_{p = 1}^P {b_p} {\frac{{p\pi }}{{{P_{mod }}}}} } \right|^2}
= \frac{{{a_1}{\beta ^2}}{E_U}}{{{N_T}{N_S}}}
\end{aligned}
\end{equation}
where
\begin{equation}\label{eq0034}
{{a}_{1}}={{\left| \sum\limits_{p=1}^{P}{\left( \frac{{{b}_{p}}p\pi }{{{P}_{mod }}} \right)} \right|}^{2}}
\end{equation}

Given that the domination of PA's nonlinear distortion comes from the third-order IMPs, we have
\begin{equation}\label{eq0027}
\begin{aligned}
& \sum \limits_{  \delta=3,5...}\Omega  \left( {{{\eta}},{\beta^{2}{{E}_{U}}}} \right)  \approx \Omega  \left( {{{3}},{\beta^{2}{{E}_{U}}}} \right)
\text{ }\text{ }\text{ }\text{ }\text{ }\text{ }\text{ }\text{ }\text{ }\text{ }\text{ }\text{ }\text{ }\text{ }\text{ }\text{ }\text{ }
 \\
& = {\left( {\frac{{{\beta ^2}{E_U}}}{{{N_T}{N_S}}}} \right)^{{3}}}{\left| {{{\sum\limits_{p = 1}^P {{b_p}\left( {\frac{{p\pi }}{{{P_{mod }}}}} \right)} }^3}} \right|^2}
 =  {\frac{{{a_3}{\beta ^6}}}{{{N^3_T}{N^3_S}}}} {E^3_U}
\end{aligned}
\end{equation}
where
\begin{equation}\label{eq0035}
{{a}_{3}}={{\left| \sum\limits_{p=1}^{P}{{{b}_{p}}{{\left( \frac{p\pi }{{{P}_{mod }}} \right)}^{3}}} \right|}^{2}}
\end{equation}

With a similar analysis, we have the clipping distortion as
\begin{equation}\label{eq0028}
\begin{aligned}
\alpha^2 D_k E_U
& = \Omega  \left(1,D_k E_U \right)\\
& = \frac{{{D_k}{E_U}}}{{{N_T}{N_S}}}{\left| {\sum\limits_{p = 1}^P {b_p} {\frac{{p\pi }}{{{P_{mod }}}}} } \right|^2}
 =\frac{{{a_1}{D_k}}}{{{N_T}{N_S}}}{E_U}
\end{aligned}
\end{equation}

Now, with (\ref{eq0013}), (\ref{eq0024}), (\ref{eq0027}), and (\ref{eq0028}), we reconsider the equation in \ref{eq0012} and obtain the receiver SNR in a polynomial form, which is
\begin{equation}\label{eq0060}
\gamma =\frac{{{a_1}{\beta ^2}\left\| H \right\|_F^2{E_U}}}{\sigma _{ch}^2 + \frac{{{a_1}{D_k}}}{{{N_T}{N_S}}}{E_U} + \sum\limits_{s=1}^{{{N}_{S}}}{\phi \left( 3,s \right)} \cdot \frac{{{a_3}{\beta ^6}}}{{N_T^3N_S^4}}   E_U^3 }
\end{equation}

\subsection{System SER}
From \cite{ref020}, it is established that the receiver SNR's probability distribution function (PDF) is given by
\begin{equation}\label{eq0030}
P_{e}(\gamma)={\left\| \mathbf{H} \right\|_{F}^{2}e^{-\left\| \mathbf{H} \right\|_{F}^{2}}}/{\gamma \cdot \Gamma(\lambda)}
\end{equation}
where
\begin{equation}
\lambda =E\left( \left\| \mathbf{H} \right\|_{F}^{2} \right)=N_{T}\cdot N_{R}
\end{equation}
and $\Gamma(z)$ represents the Gamma function, which can be expressed as
\begin{equation}
\Gamma (z)=\int_{0}^{\infty }{{{x}^{z-1}}{{e}^{-x}}dx}
\end{equation}

After being captured at the receiver antennas and processed by the spatial stream parser, received symbols are inputted into the M-QAM decision block. As has been established in reference \cite{ref021}, the relationship between the receiver SNR and the symbol estimation error probability can be expressed as
\begin{equation}\label{eq0031}
\begin{aligned}
{{P}_{w}}(\gamma)
& =\frac{4\left( \sqrt{M}-1 \right)}{\pi \sqrt{M}}{{\int\limits_{0}^{\frac{\pi }{2}}{\left( \frac{{{\sin }^{2}}\theta }{{{\sin }^{2}}\theta +{{g}_{QAM}}\gamma } \right)^{{{N}_{R}}}}}}d\theta \\
& -\frac{4}{\pi }{{\left( \frac{\sqrt{M}-1}{\sqrt{M}} \right)}^{2}}{{\int\limits_{0}^{\frac{\pi }{2}}{\left( \frac{{{\sin }^{2}}\theta }{{{\sin }^{2}}\theta +{{g}_{QAM}}\gamma } \right)^{{{N}_{R}}}}}}d\theta
\end{aligned}
\end{equation}
where
\begin{equation}
g_{QAM}=\frac{3}{2(M-1)}
\end{equation}

Hence, the SER of a nonlinear OFDM system with clipping can be written as 
\begin{equation}\label{eq0032}
\begin{aligned}
&SER(\gamma)=\int\limits_{0}^{\infty }{{{P}_{e}}(\gamma ){{P}_{w}}(\gamma )d\gamma }\\
& \approx \left( {\frac{{\sqrt M  - 1}}{{\pi \sqrt M }}} \right)\int\limits_0^{\pi /2} {\int\limits_0^\infty  {\frac{4}{\gamma }{e^{ - \left( {\frac{{3\gamma }}{{2{{\sin }^2}\theta (M - 1)}} + \left\| {\bf{H}} \right\|_F^2} \right)}}d\gamma d\theta } }\\
& = 2\frac{\sqrt{M}-1}{\pi\sqrt{M}}
{{\left( \frac{2\left( M-1 \right)\left\| \mathbf{H} \right\|_{F}^{2}}{3\gamma } \right)}^{\lambda }}
B\left( \lambda +\frac{1}{2},\frac{1}{2} \right)  \\
& \text{ }\text{ }\text{ }\text{ }\text{ }\text{ }\text{ }\text{ }
\cdot {{F}_{1}}\left( \lambda ,\lambda+\frac{1}{2},\lambda +1,-\frac{2\left( M-1 \right)\left\| \mathbf{H} \right\|_{F}^{2}}{3\gamma } \right) \\
\end{aligned}
\end{equation}
where $B\left( x,y \right)=\int_{0}^{1}{{{t}^{x-1}}{{(1-t)}^{y-1}}dt}$ is Beta function, and ${{F}_{1}}\left( a,b,c,d \right)$ denotes the well known Gaussian hypergeometric function \cite{BetaHandBook}.

At the end of this section, we would like to re-emphasize the complexity of the SNR expression and the SER expression presented in (\ref{eq0060}) and (\ref{eq0032}), respectively. The complexity of these two expressions is simplified from two aspects. First, we remove the lengthy expressions about memory effects, given that we mainly focus on the nonlinearity of the amplifier. Second, as we have discussed in Subsection A, the application of the BFPA model significantly reduces the complexity of our IMP analysis. These efforts simplify the expressions of SER and SNR, which facilitates the following analysis and optimization.

\section{SER Optimization}\label{Sec-IV}
Section III gives the SER of the studied system in (\ref{eq0032}), while this section further optimizes the SER in different scenarios.

Before technical details, we give an outline of the following three subsections and make the following notations. Subsection A assumes a constant $\eta$ and optimizes the system operating point. The optimal SNR, the optimal, system operating point, and the optimal SER performance are denoted as $\gamma^{opt}_{E_U}$, $E_U^{opt}$, and $SER^{opt}_{E_U}$, respectively. Subsection B assumes a constant $E_U$ and optimizes the signal clipping level. We denote the optimal signal clipping level, the optimal SNR, and the optimal SER as $\eta^{opt}$, $\gamma^{opt}_{\eta}$, and $SER^{opt}_{\eta}$, respectively. Subsection C undertakes the joint optimization of ${E_{U}}$ (power of the desired signal) and $\eta$ (signal clipping level). The optimal signal clipping level and the optimal PA operating point are denoted by $\eta^{g}$ and ${E^{g}_{U}}$, respectively, where the superscript $g$ represents the global optimum. The resulting optimal values for the SNR and SER are denoted as $\gamma^{g}$ and $SER^{g}$, respectively.

\subsection{Optimization of PA Operation Point}
From (\ref{eq0032}), we can derive the function for the SER with respect to the receiver SNR. The resulting expression is presented as (\ref{eq0041}) on the following page.
\begin{figure*}[hbtp]
\begin{equation}\label{eq0041}
\small
\begin{aligned}
  & \frac{\partial SER}{\partial \gamma }= -2 \frac{ \sqrt{M}-1 }{\pi \sqrt{M}} \cdot B\left( \lambda+\frac{1}{2},\frac{1}{2} \right)
 \left\{
  \begin{aligned}
  & \frac{2\lambda \left( M-1 \right)\left\| \mathbf{H} \right\|_{F}^{2}}{3{{\gamma }^{2}}}{{\left( \frac{2\left( M-1 \right)\left\| \mathbf{H} \right\|_{F}^{2}}{3\gamma } \right)}^{\lambda -1}}\\
  & \text{ }\text{ }\text{ }\text{ }\text{ } \text{ }\text{ }\text{ }\text{ }\text{ }\text{ }\text{ }\text{ }\text{ }\text{ }\text{ }\text{ }\text{ }\text{ }\text{ }\text{ }
  \cdot {{F}_{1}}\left( \lambda ,\lambda+\frac{1}{2},\lambda +1,-\frac{2\left( M-1 \right)\left\| \mathbf{H} \right\|_{F}^{2}}{3\gamma } \right) \\
 & +{{\left( \frac{2\left( M-1 \right)\left\| \mathbf{H} \right\|_{F}^{2}}{3\gamma } \right)}^{\lambda }}
 \frac{{{\partial }  }{{F}_{1}}\left( \lambda ,\lambda +\frac{\text{1}}{\text{2}},\lambda +1,-\frac{2\left( M-1 \right)\left\| \mathbf{H} \right\|_{F}^{2}}{3\gamma } \right)}{\partial \gamma } \\
\end{aligned}
 \right\} \\
\end{aligned}
\normalsize
\end{equation}
\end{figure*}

Furthermore, with(\ref{eq0041}), we have
\begin{equation}\label{eq0042}
\frac{{\partial}}{\partial \gamma}{
{{F}_{1}}\left( \lambda ,\lambda +\frac{\text{1}}{\text{2}},\lambda +1,-\frac{2\left( M-1 \right)\left\| \mathbf{H} \right\|_{F}^{2}}{3\gamma } \right)}>0
\end{equation}
Since ${\partial SER(\gamma)}/{\partial \gamma}$ is negative, we know SER is a decreasing function of $\gamma$. In other words, to obtain the minimum SER, we need to find the maximum possible SNR.

We now look at how to find $\gamma^{opt}_{E_U}$. With the expression in (\ref{eq0029}), we obtain the partial derivation function of $\gamma\left(\eta,E_U\right)$ as the respect of $E_{U}$ as
\begin{equation}\label{eq0033}
\frac{\partial \gamma }{\partial {{E}_{U}}}=\frac{{{a}_{1}}{{\beta }^{2}}\left\| \mathbf{H} \right\|_{F}^{2}\left( \sigma _{ch}^{2}-\frac{2{{a}_{3}}\Phi \beta^{6} }{{{ {{N}_{T}}^{3}{{N}_{S}}^{4} }}}E_{U}^{3} \right)}{{{\left( \frac{{{a}_{1}}{{D}_{k}}}{{{N}_{T}}{{N}_{S}}}{{E}_{U}}+\frac{{{a}_{3}}\beta^{6}\Phi }{{{ {{N}_{T}}^{3}{{N}_{S}}^{4} }}}E_{U}^{3}+\sigma _{ch}^{2} \right)}^{2}}}
\end{equation}

From (\ref{eq0033}), we know that
\begin{equation}\label{eq0038}
\frac{{\partial \gamma }}{{\partial {E_U}}}\left\{ \begin{aligned}
 > 0,{E_U} < \frac{{{N_T}N_S^{4/3}}}{{{\beta ^2}}}\sqrt[3]{{\frac{{\sigma _{ch}^2}}{{2{a_3}\Phi }}}}\\
 = 0,{E_U} = \frac{{{N_T}N_S^{4/3}}}{{{\beta ^2}}}\sqrt[3]{{\frac{{\sigma _{ch}^2}}{{2{a_3}\Phi }}}}\\
 < 0,{E_U} > \frac{{{N_T}N_S^{4/3}}}{{{\beta ^2}}}\sqrt[3]{{\frac{{\sigma _{ch}^2}}{{2{a_3}\Phi }}}}
\end{aligned} \right.
\end{equation}

As we can see from (\ref{eq0038}), $\partial\gamma/\partial E_{U}$ turns from positive to negative with the increase of $E_U$. Therefore, when $\partial\gamma/\partial E_{U}=0$, $\gamma$ reaches its maximum value $\gamma^{opt}_{E_U}$, and the corresponding $E_{U}$ can be written as
\begin{equation}\label{eq0039}
E_{U}^{opt}=\frac{{{N}_{T}}{{N}_{S}}^{\frac{4}{3}}}{\beta^{2}}\sqrt[3]{\frac{\sigma _{ch}^{2}}{2{{a}_{3}}\Phi }}
\end{equation}

Introducing  (\ref{eq0039}) into (\ref{eq0029}), we have
\begin{equation}\label{eq0040}
{\gamma_{E_U}^{opt}} = \frac{{{a_1}{\beta ^2}{N_T}N_S^{4/3}\left\| H \right\|_F^2{{\left( {2{a_3}\Phi } \right)}^{ - 1/3}}\sigma _{ch}^{2/3}}}{{{a_1}N_S^{1/3}{D_k}{{\left( {2{a_3}\Phi } \right)}^{ - 1/3}}\sigma _{ch}^{2/3} + 2{\beta ^2}\sigma _{ch}^2}}
\end{equation}

Introducing (\ref{eq0040}) to (\ref{eq0032}), the optimal SER performance can be written as
\begin{equation}\label{eq0043}
\small
\begin{aligned}
&SER^{opt}_{E_U}=2\frac{\sqrt{M}-1}{\pi\sqrt{M}}
  \cdot{{\left( \frac{2\left( M-1 \right)\left\| \mathbf{H} \right\|_{F}^{2}}{3\gamma_{E_U}^{opt}} \right)}^{\lambda }} \cdot\\
&   B( \lambda +\frac{1}{2},\frac{1}{2} )
    {{\cdot } }{{F}_{1}}\left( \lambda ,\lambda+\frac{1}{2},\lambda +1,-\frac{2\left( M-1 \right)\left\| \mathbf{H} \right\|_{F}^{2}}{3\gamma_{E_U}^{opt} } \right) \\
\end{aligned}
\normalsize
\end{equation}

\subsection{Optimization of Signal Clipping Level}
We see from (\ref{eq0003}) that the derivation function of power scaling factor $\beta$ with respect to $\eta$ should be written as
\begin{equation}\label{eq0044}
\begin{aligned}
\frac{{\partial \beta }}{{\partial \eta }} =\int\limits_\eta ^\infty  {{e^{ - {t^2}}}dt + \eta } {e^{ - {\eta ^2}}} > 0
\end{aligned}
\end{equation}

From (\ref{eq0044}), we know that the power scaling factor $\beta$ is a monotone increasing function of $\eta$. We are interested in finding the optimal $\beta$ so that we can do a simple mapping to obtain the optimal $\eta$.

The partial derivation function for $\gamma$ with respect to $\beta^{2}$ can be calculated as
\begin{equation}\label{eq0045}
\frac{\partial \gamma }{\partial {{\beta }^{2}}}=\frac{\left(\frac{a_{1}D_{k}E_{U}}{N_{T}N_{S}}+\sigma_{ch}^{2}-\frac{2a_{3}\Phi E_{U}^{3}}{{{N}_{T}}^{3}{{N}_{S}}^{4}}\beta^{6} \right){{a}_{1}}\left\| \mathbf{H} \right\|_{F}^{2}{{E}_{U}}}{{{\left( \frac{{{a}_{1}}{{D}_{k}}{{E}_{U}}}{{{N}_{T}}{{N}_{S}}}+\frac{{{a}_{3}}{{\beta }^{6}}\Phi E_{U}^{3}}{{{N}_{T}}^{3}{{N}_{S}}^{4}}+\sigma _{ch}^{2} \right)}^{2}}}
\end{equation}

Let (\ref{eq0045}) be zero. We note that the optimal power scaling factor can be calculated through the following equation:
\begin{equation}\label{eq0046}
\frac{\partial \gamma }{\partial {{\beta }^{2}}}=0\Leftrightarrow
 - \frac{{2{a_3}\Phi E_U^3}}{{N_T^3N_S^4}}{\beta ^6} + \frac{{{a_1}{D_k}{E_U}}}{{{N_T}{N_S}}} + \sigma _{ch}^2 = 0
\end{equation}

Further consider $0< \beta^2<1$ ,the only solution of (\ref{eq0046}) can be written as
\begin{equation}\label{eq0047}
\beta ^{2}=\frac{{{N}_{S}}{{N}_{T}}^{\frac{2}{3}}}{{{E}_{U}}}\sqrt[3]{\frac{\left( {{a}_{1}}{{D}_{k}}{{E}_{U}}+\sigma _{ch}^{2} \right)}{2{{a}_{3}}\Phi }}
\end{equation}

With (\ref{eq0047}), it can be proved that
\begin{equation}\label{eq0048}
\frac{{\partial \gamma }}{{\partial {\beta^2}}}\left\{ \begin{aligned}
 &> 0,0<{\beta^2} < \frac{{{N}_{S}}{{N}_{T}}^{\frac{2}{3}}}{{{E}_{U}}}\sqrt[3]{\frac{\left( {{a}_{1}}{{D}_{k}}{{E}_{U}}+\sigma _{ch}^{2} \right)}{2{{a}_{3}}\Phi }}\\
 &= 0,{\beta^2} = \frac{{{N}_{S}}{{N}_{T}}^{\frac{2}{3}}}{{{E}_{U}}}\sqrt[3]{\frac{\left( {{a}_{1}}{{D}_{k}}{{E}_{U}}+\sigma _{ch}^{2} \right)}{2{{a}_{3}}\Phi }}\\
 &< 0,{\beta^2} > \frac{{{N}_{S}}{{N}_{T}}^{\frac{2}{3}}}{{{E}_{U}}}\sqrt[3]{\frac{\left( {{a}_{1}}{{D}_{k}}{{E}_{U}}+\sigma _{ch}^{2} \right)}{2{{a}_{3}}\Phi }}
\end{aligned} \right.
\end{equation}

We see from (\ref{eq0048}) that $\partial \gamma / \partial \beta^2 $ changes from positive to negative with the growing up of $\beta^2$. As a result, there is an optimal signal clipping level ${\eta ^{opt}}$ that can result in the maximum $\gamma$. With (\ref{eq0048}) and (\ref{eq0003}),\ we can then calculate ${\eta ^{opt}}$ as
\begin{equation}\label{eq0049}
{\eta ^{opt}} = \ln \left( {\sqrt {1 - \sqrt {\frac{{{N_S}N_T^{\frac{2}{3}}}}{{{E_U}}}\sqrt[3]{{\frac{{{a_1}{D_k}{E_U} + \sigma _{ch}^2}}{{2{a_3}\Phi }}}}} } } \right)
\end{equation}

We already know from subsection A that SER is a decreasing function of receiver SNR. Hence, we substitute (\ref{eq0049}) into (\ref{eq0032}) and obtain the optimal SER as
\begin{equation}\label{eq0050}
\small
\begin{aligned}
&SER^{opt}_{\eta}=2\frac{\sqrt{M}-1}{\pi\sqrt{M}}  \cdot{{\left( \frac{2\left( M-1 \right)\left\| \mathbf{H} \right\|_{F}^{2}}{3\gamma_{\eta}^{opt}} \right)}^{\lambda }}\\
& B( \lambda +\frac{1}{2},\frac{1}{2} )
{{\cdot }\text{ } }{{F}_{1}}\left( \lambda ,\lambda+\frac{1}{2},\lambda +1,-\frac{2\left( M-1 \right)\left\| \mathbf{H} \right\|_{F}^{2}}{3\gamma_{\eta}^{opt} } \right)
\end{aligned}
\normalsize
\end{equation}

Then, we see that the optimal SNR can be written as
\begin{equation}\label{eq0051}
\gamma _\eta ^{opt}
 = \frac{{{a_1}\beta _{opt}^2\left\| {\bf{H}} \right\|_F^2{E_U}}}{{\frac{{{a_3}\beta _{opt}^6\Phi }}{{{N^3_T}N_S^4}}E_U^3 + \frac{{{a_1}{D_k}}}{{{N_T}{N_S}}}{E_U} + \sigma _{ch}^2}}
\end{equation}

\subsection{Joint Optimization}
The derivation function of $\gamma (\eta,E_U)$ is
\begin{equation}\label{eq0052}
\begin{aligned}
 &\frac{{{\partial ^2}\gamma }}{{\partial \beta \partial {E_U}}}
 = \frac{{{\partial ^2}}}{{\partial \beta \partial {E_U}}}\left( {\frac{{{a_1}{\beta ^2}\left\| \mathbf{H} \right\|_F^2{E_U}}}{{\frac{{{a_1}{D_k}{E_C}}}{{{N_T}{N_S}}} + \frac{{{a_3}{\beta ^6}\Phi E_U^3}}{{N_T^3N_S^4}} + \sigma _{ch}^2}}} \right)\\
 &\text{ }\text{ }\text{ }\text{ }\text{ }\text{ }
 =\frac{\partial }{{\partial {E_U}}}\left( {\frac{{{a_1}{{\left( {{\beta ^{g}}} \right)}^2}\left\| \mathbf{H} \right\|_F^2{E_U}}}{{\frac{{{a_1}{D_k}{E_C}}}{{{N_T}{N_S}}} + \frac{{{a_3}{{\left( {{\beta ^{g}}} \right)}^6}\Phi E_U^3}}{{N_T^3N_S^4}} + \sigma _{ch}^2}}} \right)\\
 &\text{ }\text{ }\text{ }\text{ }\text{ }\text{ }
 =\frac{{{a_1}{\left( {{\beta ^{g}}} \right)^2}\left\| {\bf{H}} \right\|_F^2\sigma _{ch}^2 - \frac{{2{a_1}{a_3}{{\left( {{\beta ^{g}}} \right)}^8}\left\| {\bf{H}} \right\|_F^2\Phi E_U^3}}{{N_T^3N_S^4}}
 }}{{{{\left( {\frac{{{a_1}{D_k}{E_C}}}{{{N_T}{N_S}}} + \frac{{{a_3}{{\left( {{\beta ^{g}}} \right)}^6}\Phi E_U^3}}{{N_T^3N_S^4}} + \sigma _{ch}^2} \right)}^2}}}
\end{aligned}
\end{equation}
where $\beta^{g}$ is the optimal signal power scaling factor that can result in $\eta^{g}$.

Substituting (\ref{eq0003}) to (\ref{eq0052}) and letting $\frac{{{\partial ^2}\gamma }}{{\partial \beta \partial {E_U}}} = 0$, we have
\begin{equation}\label{eq0054b}
E_U^{g} = \frac{{N_S^2 - \sigma _{ch}^2}}{{{a_1}{D_k}}}
\end{equation}
\begin{equation}\label{eq0054a}
{\eta ^{g}} = \ln \left( {\sqrt {1 - \sqrt {\frac{{{a_1}{D_k}{N_T}N_S^{\frac{4}{3}}\sqrt[3]{{\frac{{\sigma _{ch}^2}}{{2{a_3}\Phi }}}}}}{{N_S^2 - \sigma _{ch}^2}}} } } \right)
\end{equation}

With (\ref{eq0029}), (\ref{eq0032}), (\ref{eq0054b}) and (\ref{eq0054a}), we obtain $SER^{g}$ and $\gamma^{g}$ as (\ref{eq0055}) and (\ref{eq0056}) presented in the next page.
\begin{figure*}[htbp]
\begin{equation}\label{eq0055}
SER^{g}= \frac{2(\sqrt{M}-1)}{\pi\sqrt{M}}
 \cdot{{\left( \frac{2\left( M-1 \right)\left\| \mathbf{H} \right\|_{F}^{2}}{3\gamma^{g}} \right)}^{\lambda }}B\left( \lambda +\frac{1}{2},\frac{1}{2} \right)
 \cdot {{F}_{1}}\left( \lambda ,\lambda+\frac{1}{2},\lambda +1,-\frac{2\left( M-1 \right)\left\| \mathbf{H} \right\|_{F}^{2}}{3\gamma^{g} } \right)
\end{equation}
\end{figure*}
\begin{figure*}[htbp]
\begin{equation}\label{eq0056}
\gamma^{g}  = \frac{{{N_T}{N_S}\left\| H \right\|_F^2\Omega  \left( {1,{N_T}N_S^{4/3}\sqrt[3]{{\frac{{\sigma _{ch}^2}}{{2{a_3}\Phi }}}}} \right)}}
{\begin{array}{l}
{N_S^{-1}}\Omega  \left( {3,{N_T}N_S^{4/3}\sqrt[3]{{\frac{{\sigma _{ch}^2}}{{2{a_3}\Phi }}}}} \right)\sum\limits_{s = 1}^{{N_S}} {\phi(3,s)}
+\Omega  \left( {1,\frac{{N_S^2 - \sigma _{ch}^2}}{{{a_1}}}} \right)+\sigma _{ch}^2
\end{array}}
\end{equation}
\end{figure*}

\section{Simulations and Discussions}\label{Sec-V}
\begin{figure}[htbp]
  \centering
  \includegraphics[width=0.415\textwidth]{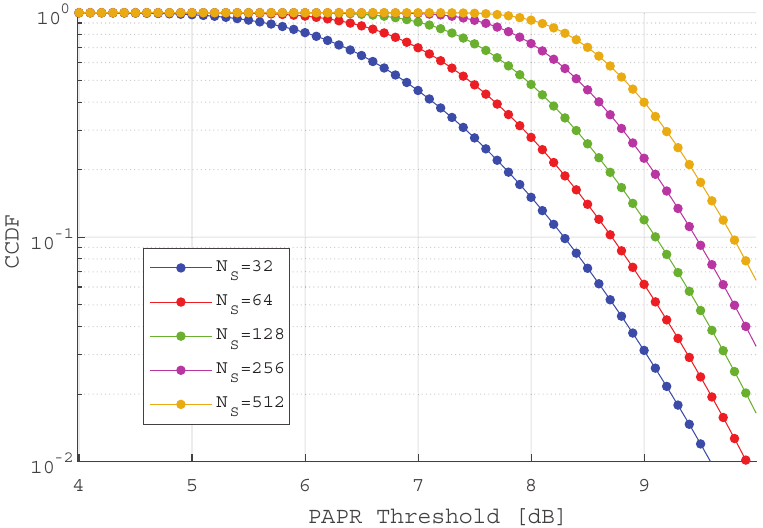}\\
  \caption{How the system's CCDF changes with PAPR under different subcarrier numbers. For detailed explanations about how CCDF is defined and calculated, we refer readers to Section III of \cite{ref064}.}\label{CCDF}
\end{figure}

At the beginning of this section, we explain why the increase in the number of subcarriers makes PA's nonlinearity a more challenging problem. Here we focus on a linear OFDM system's complementary cumulative distribution function (CCDF), which represents the probability of the system's actual PAPR exceeding a targeted PAPR threshold \cite{ref064}. As we can see from Fig. \ref{CCDF}, given the same PAPR threshold (say 9dB), OFDM systems with larger $N_S$ have significantly higher CCDFs. In this regard, with the rapid increase of $N_S$, modern wireless communication systems have much higher CCDF than before, making the OFDM system more sensitive to PA's nonlinearity. The above observation justifies the motivation of considering nonlinear amplifiers in this research.

The rest of this section validates our analytical results by simulating the studied OFDM system with signal clipping distortion and PA nonlinear distortion. We then study how the number of antennas affects the system's SER performance and examine the total degradation (TD) performance of the system.

In the following simulations, we set $N_{S}$ as 64, which corresponds to the length of the OFDM modulator/demodulator block. The over-sampling rate of the OFDM symbols is set to 4. Furthermore, unless stated otherwise, we consider a MIMO configuration with a $2\times 2$ antenna setting and employ 4-QAM as the default modulation scheme. We assume that the quasi-static Rayleigh channel is normalized, and we refer readers to \cite{ref070} for implementation details of the channel.

\begin{figure}[htbp]
  \centering
  % Requires \usepackage{graphicx}
  \includegraphics[width=0.4\textwidth]{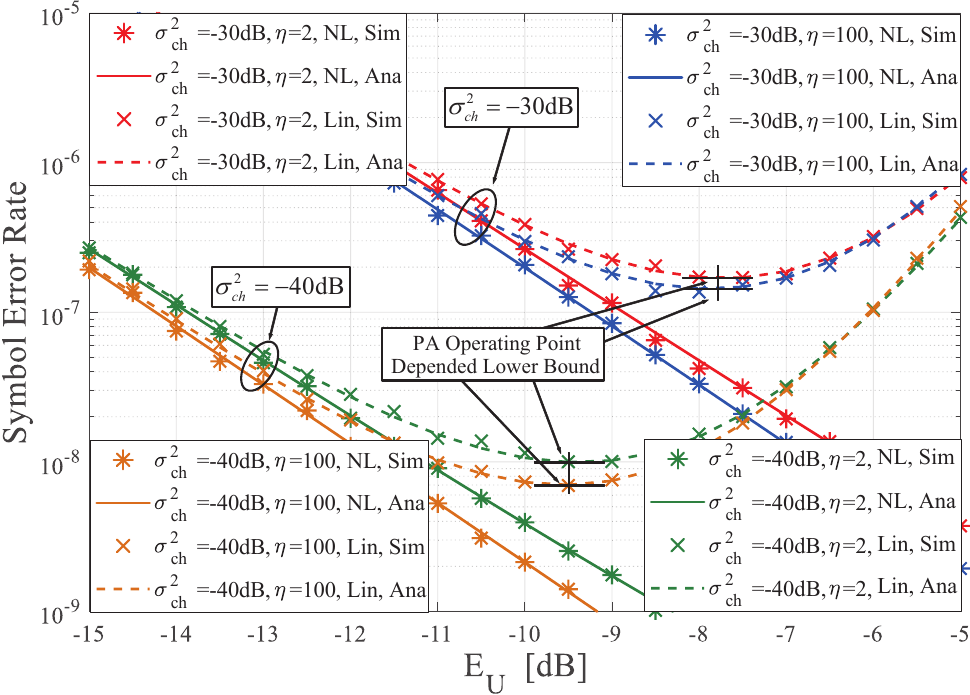}\\
  \caption{How SER changes with $E_U$ under different noise and clipping settings. Both simulated (Sim) and analytical (Ana) results consider two different assumptions: nonlinear (NL) PAs and linear (Lin) PA.}\label{Eu_SER}
\end{figure}

\begin{figure}[htbp]
  \centering
  % Requires \usepackage{graphicx}
  \includegraphics[width=0.4\textwidth]{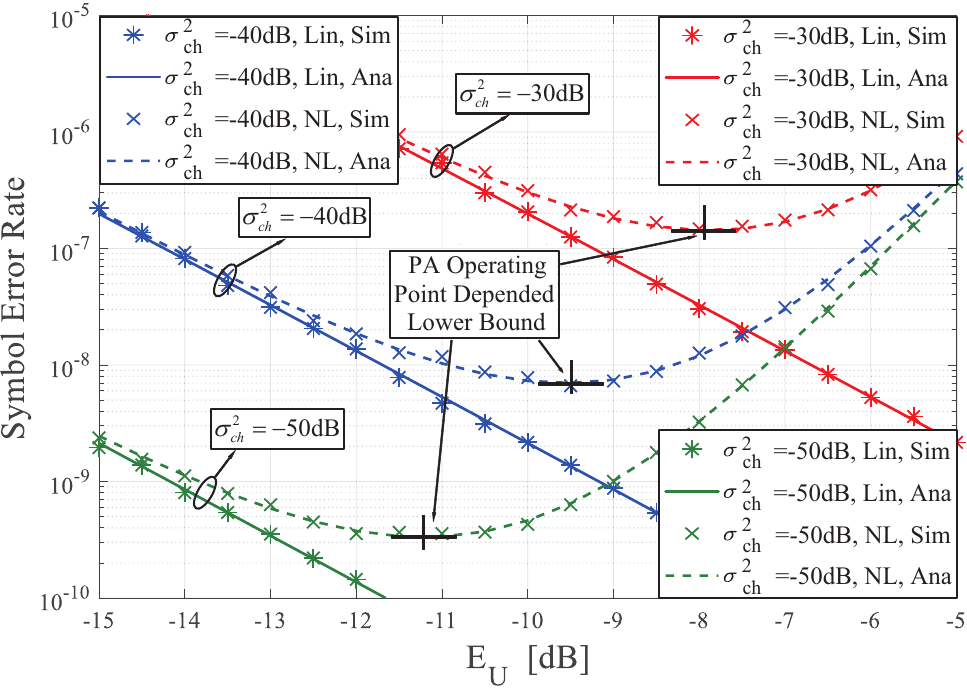}\\
  \caption{The influence of channel noise under different $E_U$. Here we let $\eta=100$. Both simulated (Sim) and analytical (Ana) results consider two different assumptions: nonlinear (NL) PAs and linear (Lin) PAs.} \label{Eu_SER_eta100}
\end{figure}

\begin{figure}[htbp]
  \centering
  % Requires \usepackage{graphicx}
  \includegraphics[width=0.39\textwidth]{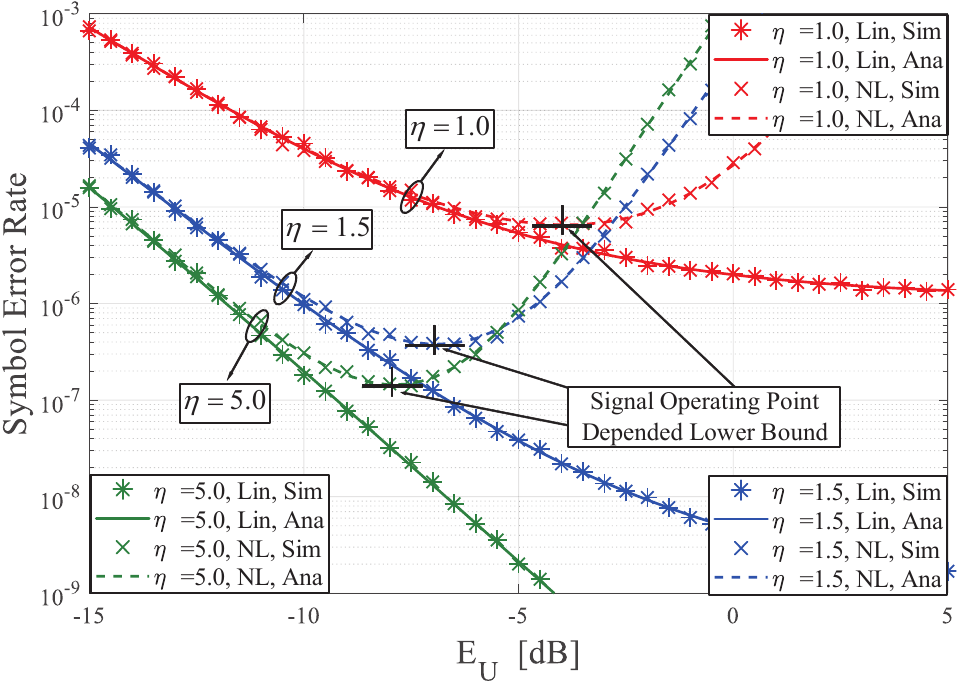}\\
  \caption{The influence of signal clipping level under different $E_U$. Here we let channel noise be $-30dB$. Both simulated (Sim) and analytical (Ana) results are presented.}\label{Eu_SER_deta}
\end{figure}

Fig. \ref{Eu_SER}, Fig. \ref{Eu_SER_eta100}, and Fig. \ref{Eu_SER_deta} are presented to examine the influence of the PA operating point $E_{U}$ on the SER performance under various system configurations while keeping $\eta$ constant. These figures provide a comprehensive understanding of the relationship between $E_{U}$ and SER, with a comparison of simulated and analytical curves.

In Fig. \ref{Eu_SER}, it is evident that each nonlinear case exhibits an SER lower bound that is dependent on $E_{U}$. This phenomenon can be explained as follows: when $E_{U}$ is not large, the PA stays in its linear region, hence its nonlinearity could be negligible. The increase of $E_{U}$ raises the PA's input power and leads to a higher SNR and a lower SER. However, when $E_{U}$ is increased beyond a certain point, further increasing it results in very high nonlinear distortion. An SER lower bound can be observed when the detrimental effect of nonlinear distortion counteracts the amplification of the desired signal. Here the PA's operating point that results in this SER lower bound is denoted as $E^{opt}_U$. Beyond $E^{opt}_U$, further increase $E_{U}$ deteriorates the SER performance.

From Fig. \ref{Eu_SER_eta100}, we can see that the lower bound of system SER and $E^{opt}_U$ is positively correlated with $\sigma_{ch}^{2}$, the channel noise. This is, a higher noise level necessitates a higher $E_{U}$ to reach the optimal SER, but the resulting SER lower bound will be larger compared to scenarios with lower noise levels. Furthermore, Fig. \ref{Eu_SER_deta} demonstrates a negative relationship between $E^{opt}_U$, the SER lower bound, and $\eta$, the clipping level.
\begin{figure}[htbp]
  \centering
  % Requires \usepackage{graphicx}
  \includegraphics[width=0.405\textwidth]{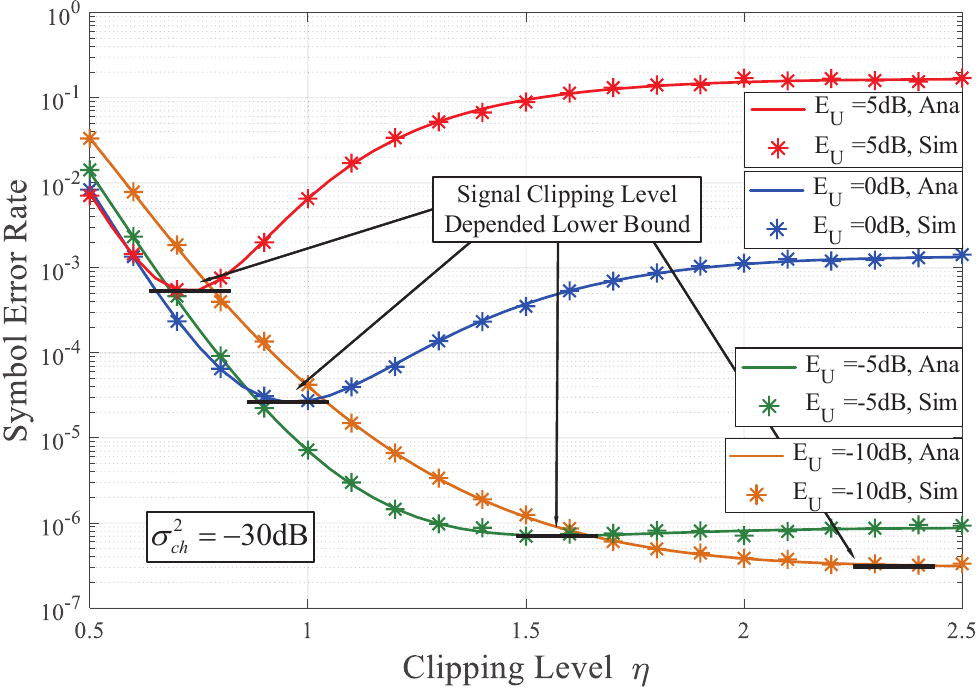}\\
  \caption{The performance of SER as a function of signal clipping level $\eta$ when nonlinear PAs are considered. Both analytical (Ana) and simulated (Sim) results are presented.}\label{eta_SER}
\end{figure}

Fig. \ref{eta_SER} investigates $\eta$'s impact on the SER when $E_{U}$ is constant. Both analytical and simulated SER curves are presented, considering the presence of significant nonlinear distortion in the PA. Notably, each curve exhibits an SER lower bound that is clipping level-dependent. When the PA operating point is sufficiently high, such as $E_{U}=5$ dB or $0$ dB, increasing the clipping level initially leads to a decrease in SER. This is because signal clipping can help mitigate the nonlinear distortion of a PA by reducing the signal's PAPR. However, further, increasing $\eta$ beyond the optimal value can result in significant degradation of the system's SER performance because clipping noise now dominates the system's impairment.
\begin{figure}[htbp]
  \centering
  \includegraphics[width=0.4\textwidth]{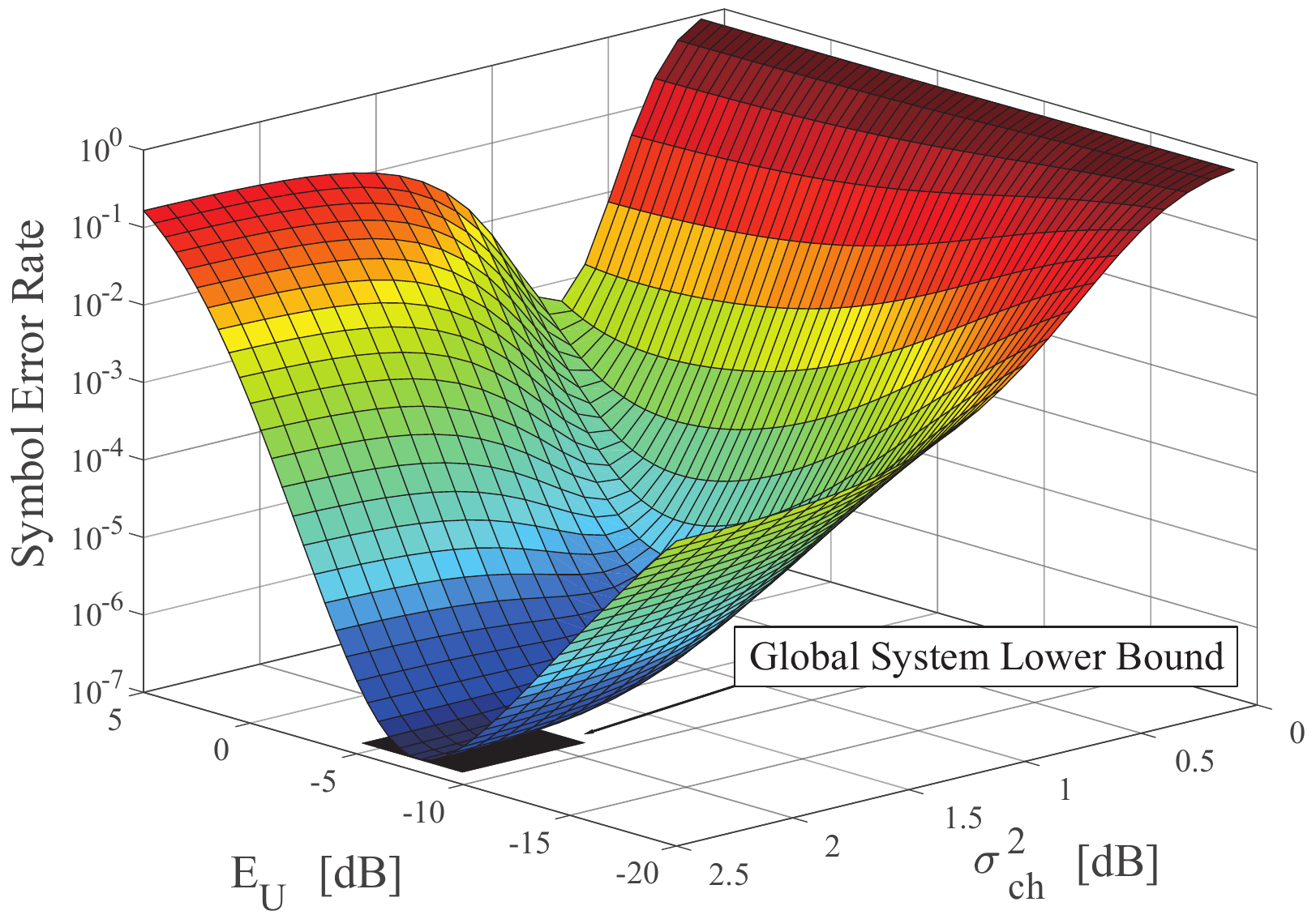}\\
  \caption{A 3D figure showing the relationship between 1) system SER, 2) signal clipping level $\eta$, and 3) PA operating point $E_{U}$. Channel noise $\sigma_{ch}^{2}=-30$dB. Nonlinear PAs are considered.}\label{eta_Eu_SER}
\end{figure}
\begin{figure}[htbp]
  \centering
  \includegraphics[width=0.39\textwidth]{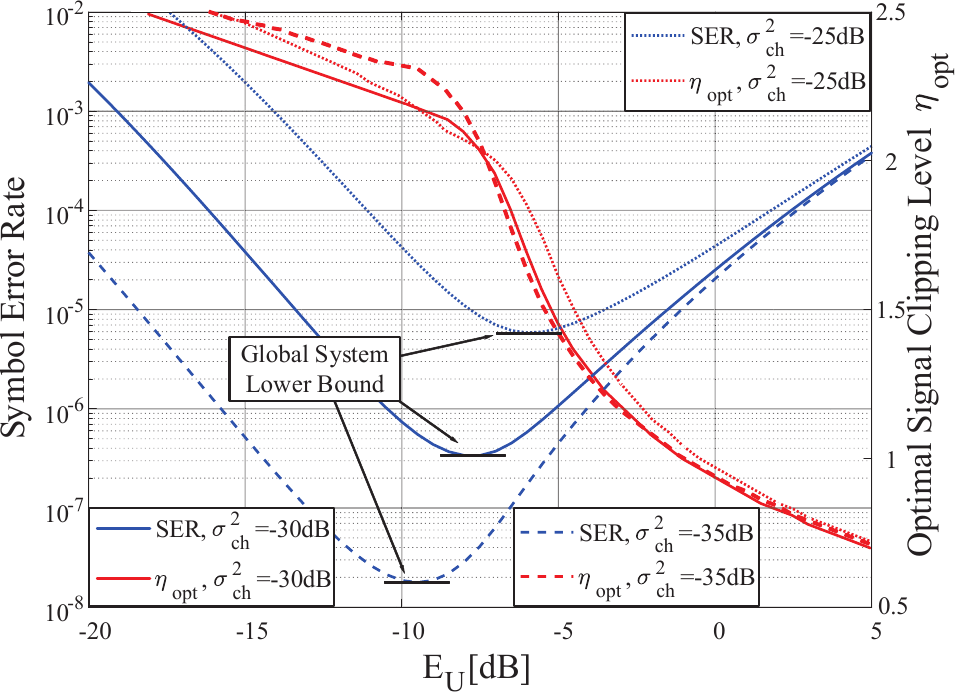}\\
  \caption{How the curves of SER performance and the curves of $\eta_{opt}$ optimal signal clipping level change according to the system operating point $E_U$.}\label{Fig_Eu_SER_etaopt}
\end{figure}

In Fig. \ref{eta_Eu_SER}, we treat SER as a function of both PA operating point $E_{U}$ and the signal clipping level $\eta$, with a consideration of nonlinear PAs. As anticipated in Section IV, we can observe a global optimal SER in the figure. Specifically, when the channel noise $\sigma_{ch}^{2}$ is set to $-30$ dB, the global optimal SER is determined to be $8.57\times 10^{-6}$, which aligns precisely with the simulated results.

Fig. \ref{Fig_Eu_SER_etaopt} presents how SER and the optimal signal clipping level change with the system operating point $E_U$ under various noise conditions. It can be seen that there is a global system lower bound of SER when system operating point $E_U$ sweeps. Moreover, the optimal signal clipping level $\eta_{opt}$ is a convex function, where its convex point corresponds to the system operating point. Besides, the global system lower bound is positively related to the channel noise $\sigma_{ch}^2$, and the global system lower bound of SER is also positively related to $\sigma_{ch}^2$. This can be explained by (\ref{eq0056}): $\gamma^g$ is positively related to the channel noise.

\begin{figure}[htbp]
  \centering
  \includegraphics[width=0.4\textwidth]{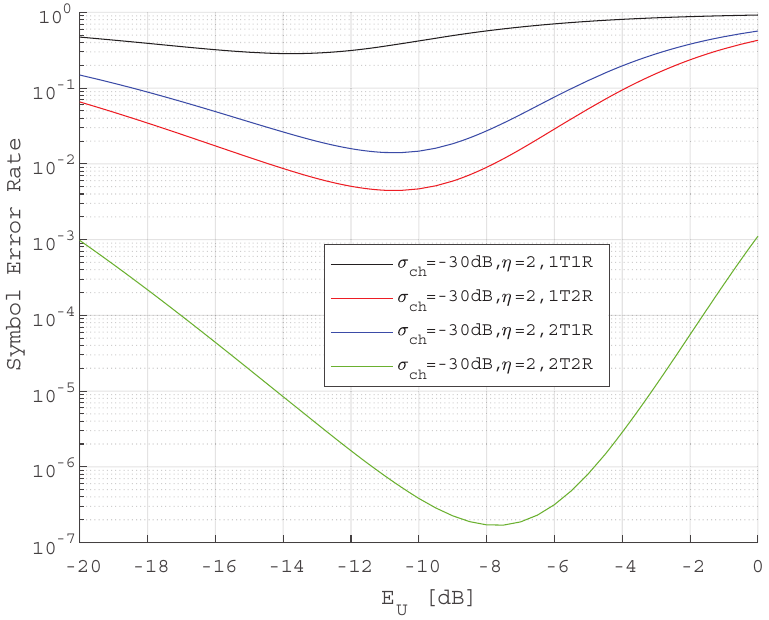}\\
  \caption{How SER changes with $E_U$ under different numbers of transmit/receive antennas. In the figure, $n$T$m$R refers to a system with $n$ transmit antennas plus $m$ receive antennas.}\label{NTNR}
\end{figure}

Fig. \ref{NTNR} presents how the number of transmit/receive antennas affects the system SER when nonlinear amplifiers are considered. As the figure shows, using more transmit/receive antennas can significantly reduce the system's SER, although it comes with increased system complexity.

\begin{figure}[htbp]
  \centering
  \includegraphics[width=0.4\textwidth]{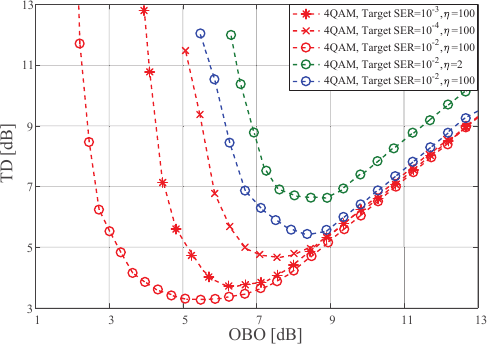}\\
  \caption{TD performance when various target SERs and different $\eta$ values are considered.
  }\label{OTO_TB_eta2}
\end{figure}

Fig. \ref{OTO_TB_eta2} presents an investigation into the relationship between the PA output back off (OBO) and the overall system TD. The OBO refers to the reduction in output power from the PA's maximum allowed output power, indicating the extent to which the PA operates in its nonlinear region. On the other hand, TD represents the SNR loss compared to that of an ideal linear amplifier in achieving a specific level of SER. For each target SER, the TD graphs in Fig. \ref{OTO_TB_eta2} exhibit a truncation point in the low-value OBO range, with the associated SER lower bound being equal to the target SER. Notably, as the PA operates within its highly nonlinear range, the TD graphs noticeably deviate from those obtained with a linear PA. It is understandable that as the target SER decreases, the OBO value at which the truncation point occurs increases. This trend becomes evident when examining the TD graph for 4-QAM signals, particularly when the target SER ranges from $10^{-2}$ to $10^{-4}$. Furthermore, a decrease in the parameter $\eta$ leads to an increase in the OBO value. This observation is apparent when analyzing the TD graph for 4-QAM signals with varying $\eta$ values from 100 to 2.

\section{Conclusion}\label{Sec-VI}
This paper studies the performance of OFDM systems with considerations of signal clipping distortion and PA nonlinearity distortion. By modeling the PA and analyzing its IMPs, we derive the system SNR and SER in a polynomial form. These derivations offer the advantage of low complexity while maintaining reasonable accuracy. Additionally, we conduct joint optimization of the clipping distortion and the PA nonlinearity to obtain the global minimum SER. This optimization process identifies the optimal signal clipping level and the optimal PA operating point. Furthermore, we examine the influence of system parameters on the SER performance under various configurations.

%\section{Acknowledgment}%The authors acknowledge the support from China National Science and Technology Major Project (Grant No. 2018ZX03001001-002). The authors acknowledge the support from the School of Electronics, Peking University, Beijing, China; the support from the Department of Information Engineering, The Chinese University of Hong Kong, Hong Kong SAR; and the support from 2012 Laboratory, Huawei Technologies Co., Ltd, Beijing, China.

\bibliographystyle{IEEEtran}
\bibliography{ref}

\end{document}